\documentclass[11pt]{article}

\usepackage{url}
\usepackage{fullpage}
\usepackage{graphicx}

\title{On the Interoperability of Decentralized\\ Exposure Notification Systems
}

\author{
       Marko Vukoli\'c\\ 
       IBM Research -- Zurich\\ 
       mvu@zurich.ibm.com
}
\date{}

\begin{document}
\maketitle

\begin{abstract}
This report summarizes the requirements and proposes a high-level solution
for interoperability across recently proposed COVID-19 exposure notification efforts. Our focus is on interoperability across exposure notification (EN) applications which are based on the decentralized Bluetooth Low Energy (BLE) protocol driven by Google/Apple Exposure Notifications API (including DP3T and similar protocols). We distinguish different interoperability use cases, such as worldwide public EN interoperability, as well as interoperability in the enterprise EN systems. This report also proposes an API and a backend implementation architecture for EN interoperability. Finally, we propose using a permissioned blockchain-based solution for managing EN backend certificates and configurations (without storing any users' data on the blockchain) for helping address EN interoperability challenges across different vendors. 
\end{abstract}

\section{Introduction}
\label{intro}

Many COVID-19 exposure notification (EN) systems have been proposed since the SARS-CoV-2 virus outbreak. Currently, the most popular proposals are decentralized exposure notification proposals based on the Google/Apple Exposure Notifications (GAEN) service using Bluetooth Low Energy (BLE) \cite{GappleBLE-EN}, capturing contacts of portable devices (e.g., smartphones), which are used to capture contacts of people (portable devices users). 

There are other exposure notification proposals which are not based on the GAEN. However, these are not the focus of this document, as the future of such systems is uncertain at this moment, as Android and iOS are expected to limit the proliferation and/or functionality of such applications in smartphones.

This report is organized as follows. 

\begin{itemize}
	\item In Section~\ref{background}, we give background overview of the GAEN BLE protocol, 
	and motivate the need for interoperability of EN solutions. 
	\item Section~\ref{sec:requirements} gives an overview of requirements for EN interoperability, refining the DP3T \cite{DP3T-interop} and EU exposure notification interoperability guidelines \cite{EU-interop}. 
	\item In Section~\ref{highlevel}, we propose a  high-level solution for interoperability, distinguishing the cases of worldwide interoperability in the global, public EN use case, as well as interoperability in the enterprise EN use case.
	\item Finally, Section~\ref{APIs} gives specific APIs for interoperability and outlines the EN interoperability implementation architecture. 
\end{itemize} 

\section{Background}
\label{background}

\subsection{Google/Apple Exposure Notification (GAEN) Protocol Overview} 

 In short, GAEN is based on Bluetooth devices periodically exchanging 16-byte Rolling Proximity Identifiers (IDs) and 4-byte Associated Encrypted Metadata (Metadata) which includes 1 byte of transmit power level used to estimate the proximity of the contact \cite{GappleBLE-EN}. IDs change every 15 minutes to make tracking of users more difficult. IDs are derived from a Temporary Exposure Key which is itself changed every 24 hours. Privacy guarantees of GAEN are out of the scope of this document.

Each device stores IDs and Metadata received over Bluetooth, corresponding to \emph{contact events}, which are later used to estimate the risk when a contact with a COVID-19 infected person is detected. The detection proceeds as follows:
\begin{itemize} 
	\item When an infected person takes a COVID-19 (positive) test, an \emph{operator} of an EN \emph{backend system} assigns a one-time code to the infected person, which allows the infected person to \emph{upload} its most recent (typically representing the last 14 days) Temporary Exposure Keys to the backend server. Temporary Exposure Keys uploaded by an infected person are referred to as Diagnosis Keys.
	\item End-user devices periodically \emph{download} Diagnosis Keys from the EN backend.
	\item Devices compute exposure risks locally and EN applications inform the user of the next steps to take.
\end{itemize}

\subsection{Public Exposure Notification}
\label{public}

As COVID-19 is a global challenge, automated public exposure notification may only help health authorities if the system works on a global scale. However, due to political and state sovereignty reasons, a single backend in an EN system will cover a relatively small geographical area. In principle, ongoing deployments of EN apps and backends are deployed in fragmented geographical regions which largely correspond to geographical borders of a sovereign country; or to states of a larger country (e.g,. United States).

\begin{figure}[htbp]
	\centering
	\includegraphics[width=0.4\linewidth]{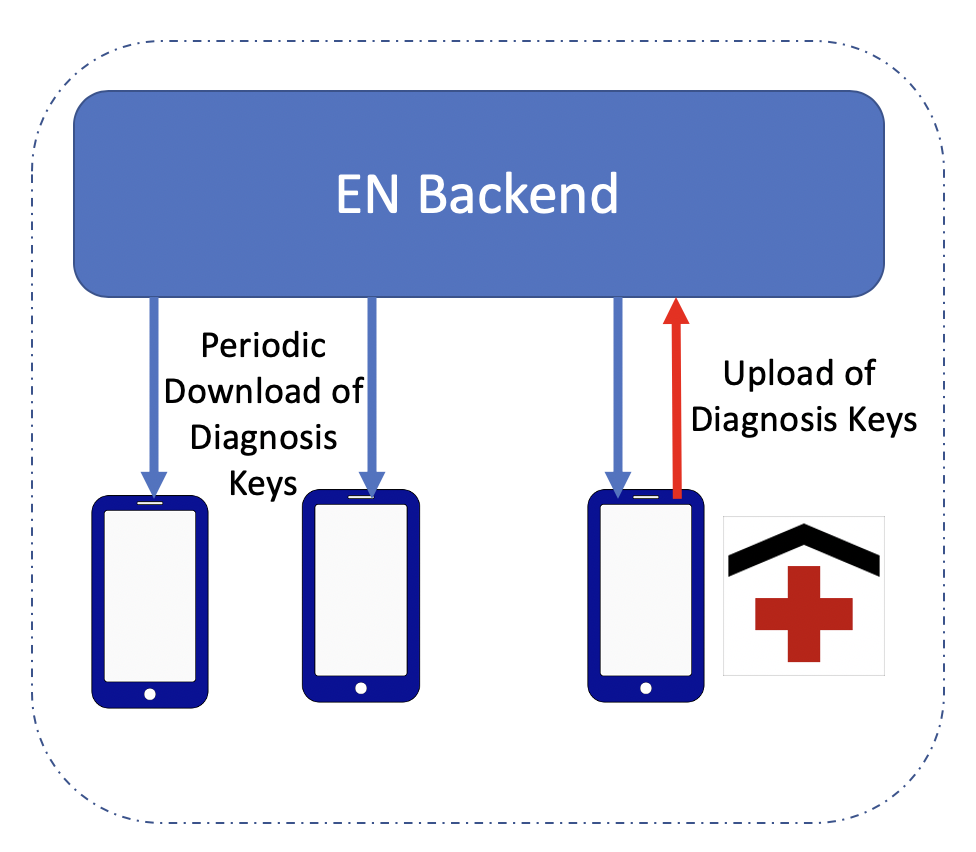}
	\caption{A single country is served by a single EN backend. Users within a country download Diagnosis Keys from this backend. Infected users upload Diagnosis Keys to the backend with permission of country health authorities.}
	\label{fig:singleregion}
\end{figure}

Hence, a relatively simple solution to exposure notification depicted in Figure~\ref{fig:singleregion} will not work with the EN system fragmented across countries, if users are ever allowed to leave their countries.  Therefore, \emph{interoperability} of GAEN-based EN systems is a key element in making EN systems usable in a global society that permits travel and movement of people. The interoperability challenges are augmented by the fact that individual governments and health authorities are choosing different operators, vendors and solutions for their EN backends. 

We foresee anywhere from two to three hundred EN backends (with division of certain large countries into state regions), which would need to collectively serve up to 8 billion users. 

\subsection{Enterprise Exposure Notification}
\label{enterpriseusecase}

In the enterprise use case, which may be deployed in different ``return-to-work'' programs and actions, we are interested in interoperability of exposure notification systems within a single enterprise. In this use case, we foresee multiple distributed enterprise \emph{sites} (which correspond to, e.g.,  different geographical branches of an enterprise) to interoperate and combine into a single logical system. In this use case, we expect up to few dozens of sites,  with at most one to two million users. In the enterprise use case, we focus on interoperability among enterprise sites, not on the interoperability with public exposure notification.

In the enterprise use case, exposure notification does not necessarily need to be based on BLE and GAEN protocols, nor implemented in smartphones. An enterprise may opt on an alternative technology, such as ultra-wideband distance measurement technologies implemented in wearables that employees might wear during working hours. Nevertheless, the interoperability considerations across different sites comprising such systems will, conceptually, remain similar to those based on interoperability within GAEN/BLE sites/countries, provided that these systems follow the similar decentralized and privacy-preserving approach as GAEN/BLE.\footnote{This document does not discuss interoperability between decentralized exposure notification systems based on different physical layer technologies.}

\section{Interoperability Requirements}
\label{sec:requirements}

\subsection{Definitions}

In the following we use a generic term \emph{region} to denote a \emph{country} in the public exposure use case (Sec.~\ref{public}), and a \emph{site} in the enterprise use case (Sec.~\ref{enterpriseusecase}).

\subsubsection{Classification of Regions}
\label{classification}

\paragraph{Mapping between regions and users --- base and roaming regions.} Each user of an EN system is mapped to one or more \emph{base} regions, corresponding to regions which the user often frequents, such as a region A covering the user's home, and regions B and C which cover the user's working places. The criteria for declaring a region as a base region should be related to epidemiological parameters. For instance, if a user Alice visits a region X at least once in 14 days, a region X could be Alice's base region.

Moreover, every user is optionally mapped to 0 or more \emph{roaming} regions. Roaming region are all the regions which do not satisfy the criteria for declaring the region as a user's base region, but which a user visits occasionally. An association between a user and a roaming region is temporary and can expire after, e.g., 14 days following the exit of a user from a roaming region. 

As we discuss later, for privacy reasons, the mapping between a user and a region should be known only to a user, to the extent possible. 

\paragraph{Mapping between regions and solution providers --- a vendor of a region.}  As we already discussed, in the public use case we expect technological solutions across regions (countries) to be different as different jurisdictions opt for different technology providers. 

To model this we introduce an attribute of a region which defines a \emph{vendor} attribute of a region. We assume that all regions with the same vendor run the same (or fully compatible) EN backend solution. 

\paragraph{Grouping of regions --- clusters.}  Regions can further be optionally grouped in \emph{clusters} which in the public use case reflect geographical and political grouping across countries. For instance, European Union (EU) countries can belong to the EU cluster, and similarly US states can belong to the USA cluster. 
A region is assumed not to belong to multiple clusters.

In the enterprise use case, we see all regions (sites) that belong to the same enterprise as a cluster.  This is not a strict requirement, yet the typical number of employees in a large enterprise allows for this assumption which simplifies the reasoning about enterprise EN interoperability (compared to public EN interoperability). 

\subsubsection{Locals and Travelers}

When two users are in contact in a given region R, we adopt the following naming convention depending on the mapping between regions and users (see Sec.~\ref{classification}).

\begin{itemize}
	\item (Local) If R is a base region of a user, a user is called a \emph{local}.
	\item (Traveler) If R is a roaming region of a user, a user is called a \emph{traveler}. 
\end{itemize}

Locals and travelers are defined with respect to a given contact event. Hence, a contact can be established between two locals, a local and a traveler, or between two travelers.

\subsection{Functional Requirements}
\label{Fs}

\begin{itemize}
	\item \textbf{F1 -- Notification to a Local about Infected Travelers.} The EN interoperability system must enable a local to receive Diagnosis Keys pertaining to infected travelers.
	
	\item \textbf{F2 -- Notification to a Traveler about Infected Locals.} The EN interoperability system must enable a traveler to receive Diagnosis Keys pertaining to infected locals.
	
	\item \textbf{F3 -- Notification to a Traveler about Infected Travelers.} The EN interoperability system must enable a traveler to receive Diagnosis Keys pertaining to infected travelers.

	\item \textbf{F4 -- Positive Test of a Roaming User.} The EN interoperability system should enable registration of Diagnosis Keys of a user diagnosed positive while roaming.\footnote{Per EU guidelines \cite{EU-interop}, roaming users should upload their relevant proximity encounter information to a base region (``home country'') backend.}

\end{itemize}

\subsubsection{Public EN Interoperability Rollout Phases} 

Complete fulfillment of Requirements F1 and F3 is dependent on Requirement F4. Partial fulfillment of Requirements F1 and F3 (covering majority of cases) is possible without F4. The public EN interoperability solution can be rolled-out in two phases: 
\begin{itemize}
	\item \textbf{Phase I:} complete fulfillment of F2 and partial fulfillment of F1 and F3, for Diagnosis Keys which pertain to positive tests and subsequent registration in users' base regions.
	\item \textbf{Phase II:} complete fulfillment of F1, F2, F3 and F4. 
\end{itemize}

Interoperability across different regions can also be expected to proceed at a different pace, depending on whether or not they belong to the same cluster, or have the same vendor. For example, at a given time, an interoperability solution can be in Phase II with respect to regions with vendor IBM, while being in Phase I with respect to some other vendors, and no interoperability with the remaining vendors. 

For global scale functionality of the public EN system, Phase II interoperability across all regions, regardless of their vendor or cluster, is a desired outcome. 

 \subsection{Scalability Requirements}
 
 \begin{itemize}
 	\item \textbf{S1 -- Global Scalability of Public EN.} The public EN interoperability system must be able to scale to hundreds of regions and billions of users globally.
 \end{itemize}

\subsection{Security and Privacy Requirements}
\label{spreq}

\begin{itemize}
	\item \textbf{SP1 -- Maintaining Single-Region Guarantees.} The EN interoperability system must retain, to the extent possible, the security and privacy provided by single region solutions. Departure from single-region security and privacy guarantees are only allowed in order to satisfy the scalability requirement (S1), and only with users' explicit consent.
	
	\item \textbf{SP2 --  Roaming Privacy.} Individual regional EN backends should learn minimum possible information about users' roaming patterns.
	
	\item \textbf{SP3 -- Deletion of Uploaded Data.} Any data uploaded by infected users to backends comprising the EN interoperability system, must be deleted by individual backends within T days of the date that such data was available at a given backend. T is defined by the respective regional health authority/government. In the absence of such a regulation, T defaults to 30 days. 
	
	\item \textbf{SP4 -- Authenticity of Data Across Jurisdictions.} The EN interoperability system deployed across different jurisdictions should be made publicly available while providing adequate security guarantees, particularly around the authenticity of test data \cite{EU-interop}.
	
	\item \textbf{SP5 -- Security of Cross-Backend Communication.} The EN interoperability system should allow individual regional backends to communicate and receive relevant Diagnosis Keys using a trusted and secure mechanism \cite{EU-interop}.
	
\end{itemize}

\subsection{Usability Requirements}
\label{usability}

\begin{itemize}
	\item \textbf{U1 -- Minimizing the number of EN Apps.}  Users should be able to rely on a single EN app independently of the region they are in at a given moment. This must be the case for regions of the same vendor. If this is not possible, across regions of different vendors, the number of different EN apps across such regions must be minimized.

\item \textbf{U2 -- User Bandwidth.} Users must have at least the minimal bandwidth necessary to fulfill functional requirements of an EN interoperability system.  

\item \textbf{U3 -- User Interactions and Travel Patterns.} The number of a user's interactions with an app should be made reasonably small, especially in respect to granularity and declaration of roaming regions. 
\end{itemize}

\subsection{The tradeoff between usability/privacy and scalability requirements}
\label{tradeoff}

Requirement U3 and SP2 are, in principle, in conflict with Requirement S1. Namely, there is a tradeoff between: (1) number of users' interactions with an app needed to faithfully input all valid roaming regions and privacy of visited regions and (2) bandwidth that backends/users  need to support.\footnote{In the absence of any regional information provided by users, the only way to satisfy F1 is to feed into a region R all Diagnosis Keys from all regions from which users are permitted to roam into region R (all-to-all replication).}

One way to approach this tradeoff, is to favor Requirement U3 (and SP2) for regions belonging to a cluster (essentially treating a cluster as one big region), whereas S1 can be favored across clusters.

\section{A High-Level EN Interoperability Solution}
\label{highlevel}

In this section, we first propose a high-level solution for EN interoperability, focusing on satisfying individual functional requirements (see Sec.~\ref{Fs}) in the context of global, worldwide public EN interoperability. Later, in Section~\ref{enterprise}, we discuss the specific case of interoperability across EN backends belonging to the same enterprise (i.e., enterprise EN use case).

\subsection{Satisfying Requirement F1}
\label{F1}

\noindent \textbf{F1 -- Notification to a Local about Infected Travelers.} The EN interoperability system must enable a local to receive Diagnosis Keys pertaining to infected travelers.\\

\begin{figure}[htbp]
	\centering
	\includegraphics[width=0.6\linewidth]{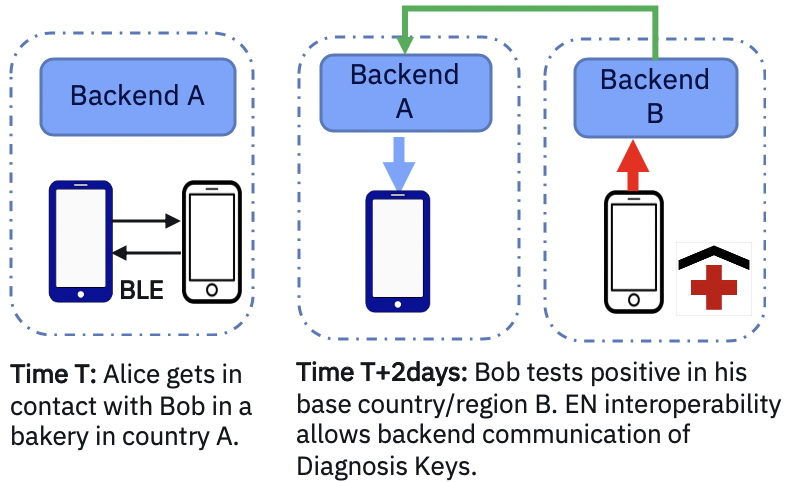}
	\caption{Satisfying Requirement F1. Backend-to-backend replication is needed. Diagnosis keys should be partially replicated, from backend B to backend A, on a need-to-know basis, unless this impairs usability/privacy (all-to-all replication might be preferred to partial replication across backends belonging to the same cluster). In this example, Bob reveals his travel patterns to his base region (home country) B backend, which makes Bob's Diagnosis Keys available to country A backend, which, in turn, notifies Alice.}
	\label{fig:f1}
\end{figure}

\emph{User story (see also Fig.~\ref{fig:f1}): Alice, who lives in country A, goes to a local bakery in her neighborhood. Next to her, there is a traveler, Bob, who exhibits COVID-19 symptoms and who could be from any country in the world. Bob goes back to his country B two days later and tests positive. Alice, aware of Bob's symptoms, is concerned she might be infected, too. How does Alice learn about Bob's infection, without listening to all EN backends, i.e., for Diagnosis Keys from all countries in the world?}

\subsubsection{Back-of-the-Envelope: Notifying All Users of All Diagnosis Keys}
\label{backenvelope}

As we discussed in Section~\ref{tradeoff}, in the absence of base/roaming region information uploaded by infected users to a backend, notifying all users of all Diagnosis Keys is required for satisfying Requirement F1. 

A rough estimate of the bandwidth needed by each individual user, to download \textbf{all} Diagnosis Keys pertaining to all infections worldwide is: 
\begin{center}
	14 keys * 16 bytes/key * 200'000 estimated max number of COVID-19 daily infections worldwide\footnote{We based the estimation of the maximum number of COVID-19 daily infections on the statistics available to date \url{https://www.worldometers.info/coronavirus/}.}  $\approx$ 45 MBytes of raw data per user per day.
\end{center}

As a result, a collection of backends serving the number of users equal to the entire population of United States (330 million) needs to serve roughly 16 Petabytes of data per day ($\approx 1.6 Tbps$ sustained bandwidth).\footnote{These numbers could be adjusted further for transport and metadata overheads (which increase the estimate) and the population without devices (which decreases the estimate), but for a rough calculation we will assume that these largely cancel each other out.}

These numbers suggest that if infected users upload some coarse-grained regional information to backends, this information would be very helpful to improve scalability. For usability and privacy reasons, the information about regions should not be fine grained. We recommend the granularity of users' input to be at a country level, or even more coarse grained (i.e., cluster level). To summarize, we recommend:
\begin{itemize}
	\item Either all-to-all or partial replication, on a need-to-know basis, across regions belonging to the same cluster. In all-to-all replication, all Diagnosis Keys originating in a cluster are replicated to all backends and users in that cluster. Smaller clusters, such as enterprise clusters, should opt for all-to-all replication.
	Larger clusters (like EU or USA), may want to opt for partial replication of Diagnosis Keys, in case all-to-all in-cluster replication is deemed expensive. 
	\item Partial replication, on a need-to-know basis, across worldwide regions not belonging to the same cluster.
\end{itemize}

\subsubsection{Partial Replication on Need-to-Know Basis}
\label{sec:partial}

Partial replication relies on an infected user uploading information about their base and roaming regions (Section~\ref{classification}) along with Diagnosis Keys. Denote the set of base and roaming regions that the infected user declared by $S$ and the region where the infected user was tested positive as $B$. 

Then, backend $B$ replicates the user's Diagnosis Keys to the backend of every region $R\in S\setminus\{B\}$ (and only to those regions). Only Diagnosis Keys are replicated to these regions, regional information uploaded by the user should never leave backend $B$.

Partial replication among backends should rely on two-way authentication. Partial replication can be push-based and pull-based. We recommend using with pull-based partial replication in which backends periodically pull data from other backends (other reference GAEN servers, including those from Google \cite{Google-GAEN-server} and DP3T also follow pull-based replication).

\subsubsection{All-to-All Replication Across Regions in a Cluster}
\label{sec:a2a}

Conceptually, all-to-all replication across backends across regions in one cluster can be implemented in several ways: pull-based, push-based, using a central cluster gateway (proposed as a guideline in the case of the EU cluster  \cite{EU-interop-gateway}), or not. As all-to-all replication can be seen as a special case of partial replication, which an interoperable EN backend anyway needs to implement, we recommend using the  pull-based replication similar to the one described for partial replication.

\subsubsection{Backend Certificate and Configuration Management in the Public Use Case}

With hundreds of regions/clusters, management of backend public keys/certificates should be handled automatically, and manual certificate management and roll over (as well as that of other basic backend information such as backend URLs and configurations) should be avoided.

For worldwide interoperability, we suggest a CA tree with, for instance, a World Health Organization (WHO) trusted root certificate. The root key can be then be used to  sign/issue country/cluster level certificates.  

Moreover, we recommend a (strongly or eventually) consistent decentralized service to maintain basic information about backend addresses, configurations and certificates. Possible technology to use here, in particular in the public worldwide use case, could be Hyperledger Fabric permissioned blockchain (decentralized ledger) \cite{AndroulakiBBCCC18}. Permissioned blockchain would be used here only as a decentralized repository of certificates and backend configurations (without storing any COVID-19 test results or GAEN Diagnosis Keys).

\subsection{Satisfying Requirement F2}
\label{F2}

\noindent\textbf{F2 -- Notification to a Traveler about Infected Locals.} The EN interoperability system must enable a traveler to receive Diagnosis Keys pertaining to infected locals.\\

\begin{figure}[htbp]
	\centering
	\includegraphics[width=0.6\linewidth]{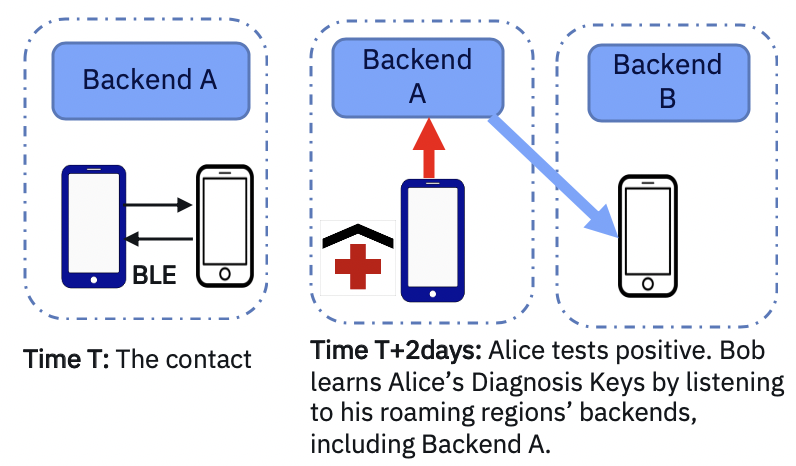}
	\caption{Satisfying Requirement F2. An EN mobile app should be able to listen to any EN  backend. A user selects which regions they want to listen to (user's base and roaming regions). }
	\label{fig:f2}
\end{figure}

\emph{User story (see also Fig.~\ref{fig:f2}): How does Bob, after returning to his country B, learn about possible infections he might have got from locals, including Alice, while in country A?}

\textbf{Base solution: Bob's app listens to roaming backend feeds (incl. country A).} Satisfying Requirement F2 is conceptually much simpler than satisfying Requirement F1. In principle, all the user needs to do is to listen to the Diagnosis Keys feeds of all its roaming regions during the visit and for 14 days after leaving a given roaming region.\footnote{This is in addition to continuous listening to feeds from all of its base regions.}

In the public EN use case, in absence of common Diagnosis Keys feed APIs across regions of different vendors, this may require updating an EN app with the roaming region public feed format and API. For this reason we discuss two alternative approaches to satisfying Requirement F2, pointing out the tradeoffs they incur.

\textbf{Alternative 1: Bob installs several apps.} To circumvent the need for an EN app from a single vendor to understand public feed formats and APIs from other vendors, a user can install several EN apps from different operators in user's roaming regions. This is at odds with Requirement U1, Sec.~\ref{usability}, and as such it is  inferior to the base solution we proposed above.

\textbf{Alternative 2: Bob listens only to his home backend B, which fetches Diagnosis Keys from backend A.} This alternative is viable for clusters which already implement all-to-all backend replication to satisfy Requirement F1 (Sec.~\ref{sec:a2a}). 

Namely, Alternative 2 practically implies all-to-all replication across backends. To see this, note that Diagnosis Keys uploaded by Alice cannot be related in any way to backend B. Hence, to notify Bob of Alice's infection, backend B needs to pull \emph{all} Diagnosis Keys from backend A. Similarly, backend B needs to pull all Diagnosis Keys from any region to which even a single user of backend B has traveled to in the last 14 days. In clusters such as EU or USA, with intensive travel patterns, this practically implies the need for all-to-all replication among backends, all the time. 

In order to satisfy security and privacy requirements SP1 and SP2 (Sec.~\ref{spreq}) backend B needs to serve such diagnosis keys to \emph{all} its users, not only to Bob. Otherwise, if backend B serves these keys only on-demand by a user, i.e., if Bob needs to select originating regions it wants to listen to, backend B can learn more than necessary about Bob's roaming patterns,\footnote{An average user should not be expected to obfuscate their roaming regions by selecting more regions than the user has actually traveled to.} which is conflict with Requirement SP2 (in contrast, in base solution for Requirement F2, no single backend alone learns more than one Bob's roaming region, satisfying SP2). Moreover, backend B would violate in this case Requirement SP1 as well, since the backend would need to `tag', in their public feeds, Diagnosis Keys based on their `origin' regions. 

\subsection{Satisfying Requirement F3}
\label{F3}

\noindent  \textbf{F3 -- Notification of a Traveler about Infected Travelers.} The EN interoperability system must enable a traveler to receive Diagnosis Keys pertaining to infected travelers.\\

\emph{User story: How does Bob, after returning to his country B from country A, learn about possible infections he might have got from other travelers while in country A?}\\

Solutions described in Section~\ref{F1} and Section~\ref{F2}, combined, satisfy Requirement F3.

\subsection{Satisfying Requirement F4}
\label{F4}

\noindent\textbf{F4 -- Positive Test of a Roaming User.} The EN interoperability system should enable registration of Diagnosis Keys of a user diagnosed positive while roaming.\\

The user app is normally capable of communicating with a base region backend, not a roaming backend. Hence, roaming users should upload their relevant proximity encounter information to a base region (``home country'') backend \cite{EU-interop}.

This requirement can be satisfied leveraging a W3C Verifiable Credentials standard \cite{W3C-VC} compatible, blockchain-based (Hyperledger Fabric) solution. 
In this solution, only 
public certificates of health certificate issuers are stored on the blockchain (with no information pertaining to users being stored on the blockchain). The details are  out of the scope of this report (an interested reader is invited to contact the author).

\subsection{Interoperability Across EN Backends within a Single Enterprise} 
\label{enterprise}

Given the sizes of even the largest enterprises, all-to-all replication, with dissemination of all Diagnosis Keys to all users, is a viable option in the enterprise EN use case. Indeed, notifying every user of all Diagnosis Keys within an enterprise is simpler to reason about compared to partial replication based on users'  regional input. This can be achieved either by:
\begin{itemize} 
	\item Having all-to-all cross-backend replication (see also Sec.~\ref{sec:a2a}), or 
	\item Having user's continuously receive updates from all enterprise EN backends (see also Sec.~\ref{F2}).
\end{itemize}

Any of the two approaches are sufficient on their own and are sufficient to satisfy functional requirements F1, F2 and F3. 

Should an enterprise wish to restrict the distribution of Diagnosis Keys, a partial-replication solution (Sec.~\ref{sec:partial}) similar to the public interoperability solution is a viable option. 

\section{Interoperability APIs and Implementation Architecture}
\label{APIs}

In this section, we overview our proposed EN interoperability APIs and implementation architecture.  These are inspired by and aligned, to the extent possible,  to the DP3T interoperability specification \cite{DP3T-interop-spec}, with the difference that our APIs and implementation architecture support for both partial replication and all-to-all replication (see Section~\ref{highlevel}). 

In a nutshell, an interoperable EN backend and user app need to satisfy the following requirements, which stem from our high-level solution of Section~\ref{highlevel}:
\begin{itemize}
	\item Implement \emph{backend feeds}, allowing all-to-all and partial replication APIs among backends of different regions (this is the key to satisfying Requirement F1).
	\item Allow an EN app to listen to public Diagnosis Keys feeds from multiple EN backends (this is the key to satisfying Requirement F2 in the global, public EN use case). 
\end{itemize}

In the following, we detail the backend feed APIs and then discuss specific implementation details.

\subsection{Backend Feed APIs}

Backend feeds for partial (Sec.~\ref{sec:partial}) and all-to-all (Sec.~\ref{sec:a2a}) replication of Diagnosis Keys have the same format as public (smartphone) feeds with the difference that they require authorization of downloads via mutual TLS authentication with other backends.

In the following, in the context of pair-wise communication between two backends, we distinguish a \emph{Consumer} and a \emph{Producer} backend. A Consumer backend is the backend which periodically pulls Diagnosis keys from a Producer backend. For example, in Figure~\ref{fig:f1}, backend A is a Consumer backend, whereas backend B is a Producer backend. We assume that the Consumer knows the URL of the Producer, and that it associates this Producer/URL to a given region. 

\subsubsection{All-to-All Replication APIs}
\label{a2a-API}

Recall that all-to-all replication is a viable approach for EN interoperability in clusters, in particular within a single enterprise. Assume that a public API of the backend  is implemented as  \textsc{get} $/v1/keys/\{chunk-num\}$, where $\{chunk-num\}$ is the next chunk of data for the client.
Here, the first request may be to \textsc{get} $/v1/keys$ to get all the keys (or the oldest available batch).

While a Consumer backend might reuse the public (smartphone) API of the Producer backend, to implement all-to-all-replication, and vice versa, this is not necessarily ideal, since the Consumer would end up pulling its own Diagnosis Keys. This would require additional logic in an EN backend, needed to implement Diagnosis Keys deduplication.

An alternative approach to using public URLs would be to mirror \emph{local} Diagnosis Keys, defined as Diagnosis Keys \emph{which originate at the producer backend} (and only those Diagnosis Keys) to a specific namespace (e.g., \textsc{get} $/v1/a2a/keys/\{chunk-num\}$) specifically for backend all-to-all consumers. 

This approach avoids the need for deduplication (or for introduction of regional information in the public feeds) which would arise in the case that backends would listen to public (client) feeds. Deduplication would be needed only if we assume the possibility of malicious backends --- this, however, seems out of the threat model for public EN clusters and single-enterprise use cases, for which all-to-all replication is relevant.

After pulling the Diagnosis keys from the Producer, the Consumer backend adds the consumed keys to its ingress path and eventually serves them to its clients (smarthones) using its public Diagnosis Keys feed.

Note that unlike \textsc{post} ingress Diagnosis Keys \cite{GappleBLE-EN}, Diagnosis Keys replicated among backends come without regional information.

\subsubsection{Partial Replication APIs}
\label{partial-API}

 Recall that partial replication is a basically the only practical approach to worldwide public EN interoperability (Sec.~\ref{sec:partial}).
 
 \paragraph{Producer creation of per-region feeds.} For partial replication it is essential that a consumer from region A only pulls those Diagnosis Keys which come via \textsc{post} interface of a producer that carries $A$ in the region field. 
 
 Therefore, every producer creates a separate per-consumer (per-region) feed with the following interface:
 
 \begin{itemize}
 	\item \textsc{get} $/v1/\{region-id\}/keys/\{chunk-num\}$, where $\{region-id\}$ is the Google/Apple two letter region ID of the consumer to which this feed is dedicated, and where $\{chunk-num\}$ is the next chunk of data for the client (consumer).
 	\item here, the first request may be to \textsc{get} $/v1/\{region-id\}/keys$ to get all the keys (or the oldest available batch) for the particular region ID.
 \end{itemize}

 \paragraph{Consumer pulling a per-region feed.} Otherwise, the consumer pulls the per-region feed it is interested in. This is done in a similar way to pulling an  all-to-all replication (i.e., periodically, while maintaining state related to the last pulled chunk). The processing of per-region feed at the consumer is also identical to the processing of an all-to-all feed.

 \subsection{Producer Architecture}
 
\begin{figure}[htbp]
 	\centering
 	\includegraphics[width=0.5\linewidth]{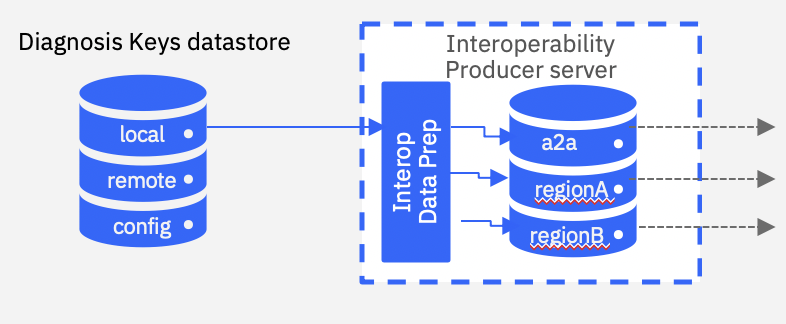}
 	\caption{Backend Architecture. Interoperability Producer  takes Diagnosis Keys from the `local' part of a Diagnosis Keys datastore and populates the backend feeds.}
 	\label{fig:producer}
 \end{figure}

 Producer architecture involves reusing and making minor modifications to a backend component responsible for creating public (smartphone) Diagnosis Keys feed, which any EN backend needs to implement (see also Fig.~\ref{fig:producer}).
 
 In the following, we assume that an EN backend distinguishes between \emph{local} and \emph{remote} Diagnosis Keys. `Local' Diagnosis Keys enter backend directly, via \textsc{post} upload of Diagnosis Keys by an infected user. On the other hand, `remote' Diagnosis Keys enter the backend via interoperability backend feeds. 
 
 A Producer's data preparation (Data Prep) engine reads \emph{only} from the `local' part of the Diagnosis Keys  datastore. 
 \begin{itemize}
 	\item For all-to-all replication,  Producer's Data Prep engine takes all keys  from the`local' part of the Diagnosis Keys datastore and prepares them under \textsc{get} $/v1/a2a/keys/\{chunk-num\}$. This is done in the same way the backeng would prepare keys for the public (smartphone) feed if there was no interoperability. 
 	\item For partial replication,  Producer's Data Prep Engine takes keys from the `local' part of the Diagnosis Keys datastore, filters them based on the regional information `local' Diagnosis Keys contain, and routes those keys under \textsc{get} $/v1/\{region-id\}/keys/\{chunk-num\}$.
 \end{itemize}
 
 As we describe next, the Consumer part of the backend populates the `remote' part of the Diagnosis Keys datastore. These keys are served by the backend to the public feed  (smartphone) \textsc{get} API \emph{only}, they are \emph{not} served by the Producer's Data Prep Engine to the backend feeds.
 
 \subsection{Consumer Architecture}
 
\begin{figure}[htbp]
 	\centering
 	\includegraphics[width=0.6\linewidth]{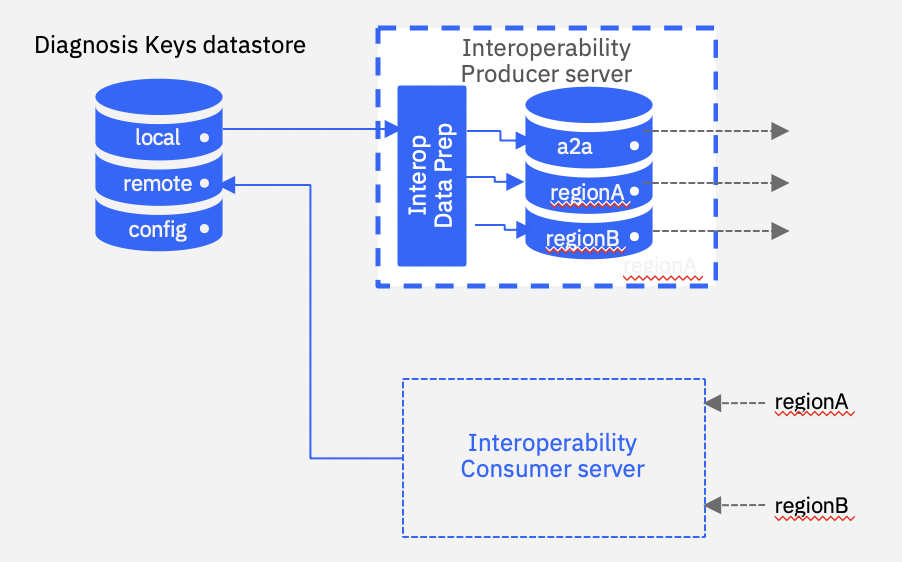}
 	\caption{Backend Architecture. Interoperability Consumer server populates the`remote' part of the Diagnosis Keys datastore.}
 	\label{fig:consumer}
 \end{figure}

 A backend of a given region runs the interoperability Consumer, which periodically pulls data from its peer (Producer) backend feeds and stores retrieved keys into the`remote' part of Diagnosis Keys datastore (Fig.~\ref{fig:consumer}). 
 
 For each producer peer (region), Consumer must maintains the following configuration:
 \begin{itemize}
 	\item $RegionID$, the two-letter region code associated with this remote region; 
 	\item $RemoteRegionURL$, The URL of the feed of the remote region. For example: $https://example.com/example-name/$;
 	\item $ReplicationType$, which denotes all-to-all or partial replication;
 	\item $Format$, the remote region format (for interoperability across vendors);
 	\item $VerificationKeys$, (optional) to verify the producer's signature on batches;
 	\item $TLSConsumerCertificate$, the TLS certificate that the consumer must use to authenticate itself to the producer. 
 \end{itemize}
 
 In addition, for each producer peer (region), Consumer must maintain the following state \cite{DP3T-interop-spec}:
 \begin{itemize}
 	\item $lastBatchID$, the ID of the last batch retrieved from this producer/region;
 	\item $recommendedNextPollTime$, (optional) the time in Linux epoch seconds,
 	at which next query should be performed. 
 \end{itemize}
 
 \subsection{Certificate Management and Access Control}
 
Producers maintain a list of Consumer certificates or corresponding CA(s) that authenticate Consumers that are allowed to query the Producer's backend feeds. In the first iteration, the management of these certificates will be manual (by backend/Producer admin). In the following iterations we propose to explore a blockchain-based (Hyperledger Fabric) solution for certificate management, in particular for worldwide public EN interoperability.

A Producer may optionally restrict access of a consumer to a certain feed only (e.g., restricting the consumer from region A only to \textsc{get}  $/v1/\{regionA\}/keys/$). 

Finally, all Producers backend feeds should be digitally signed by the same public key that a Producer uses for its public feed.

\section*{Acknowledgements}

The author is grateful to Luca De Feo, Dennis Jackson, Dennis Potashnik, Navaneeth Rameshan, Martin Schmatz, Marc Stoecklin and Carmela Troncoso, for useful discussions.\\

\bibliographystyle{abbrv}
\bibliography{fault_tolerance,blockchain,contacttracing}

\end{document}